\documentclass[showpacs,twocolumn]{revtex4}
\usepackage{amsmath}
\usepackage{amssymb}
\usepackage{graphicx}
\usepackage{epsfig}

\begin{document}

\title{Fidelity, entanglement, and information complementarity relation}
\author{Jian-Ming Cai}
\email{jmcai@mail.ustc.edu.cn}
\author{Zheng-Wei Zhou}
\email{zwzhou@ustc.edu.cn}
\author{Guang-Can Guo}
\email{gcguo@ustc.edu.cn}
\affiliation{Key Laboratory of Quantum
Information, University of Science and Technology of China,
Chinese Academy of Sciences, Hefei, Anhui 230026, China}

\begin{abstract}
We investigate the dynamics of information in isolated multi-qubit
systems. It is shown that information is in not only local form
but also nonlocal form. We apply a measure of local information
based on fidelity, and demonstrate that nonlocal information can
be directly related to some appropriate well defined entanglement
measures. Under general unitary transformations, local and
nonlocal information will exhibit unambiguous complementary
behavior with the total information conserved.
\end{abstract}

\maketitle

\section{Introduction}

Understanding the physics of quantum many-body systems is a fundamental goal
of condensed matter theory. People are interested in the study of strongly
correlated quantum states which will exhibit lots of fascinating phenomena,
such as quantum phase transitions \cite{Sachdev,Entanglement and Phase
Transitions,Osborne032110} and Kondo effect \cite{Kondo}. In the field of
quantum information processing, quantum many-body systems are also the
essential ingredient. Perspective scalable quantum computation \cite%
{Briegel2001,QC1,QC2,QC3,QC4} and communication \cite%
{Bose2003&Christandl2004} schemes based on different many-body models of
condensed matter physics have have attracted intensive interest. Moreover,
unlike the situation of bipartite entanglement, the picture of multipartite
entanglement in quantum many-body systems is still not very clear \cite%
{Plenio,Lunkes,SLOCC}.

However, the fact that the number of parameters required to describe a
quantum state of many particles grows exponentially with the number of
particles leaves a practical obstacle to the study of quantum many-body
systems. Therefore most of research focuses on the static properties of the
ground states of certain types of many-body models with quantum Monte Carlo
calculations \cite{Suzuki} and the density matrix renormalization group \cite%
{Rommer}. In this paper, we investigate the dynamics of two- and
three-qubit systems from the viewpoint of quantum information.
Through simple examples, we demonstrate perfect complementary
behavior between local information and entanglement, and reveal
the heuristic connection between measures of entanglement and
nonlocal information, which are expected to be generalized to
many-qubit systems.

Given a system of $N$ qubits, the system initial state is $|\psi\rangle_{12
\cdots N}$, with $\rho_{i}=Tr_{{1,\cdots,i-1,i+1,\cdots,n}%
}(|\psi\rangle\langle\psi|)$ the reduced density matrix of each individual
qubit. It is known that information contained in multi-qubit systems is in
two forms \cite{Horodecki}. One is local form, that is the information
content in each individual qubit $\rho_{i}$. The other is nonlocal form,
that is entanglement between different qubits. If the multi-qubit system is
isolated, i.e. initial pure and under unitary transformations only, then
information contained in the system will be transferred between different
qubits and converted between local and nonlocal forms with the total
information conserved, which results in the information dynamics \cite%
{Horodecki,peng,Teisser}. In this paper, we adopt a measure of
local information based on fidelity and reveal a heuristic
connection between measures of entanglement and nonlocal
information. Therefore, we establish an elegant complementarity
relation between local and nonlocal information. Our results link
the local information and measures of entanglement, particularly
genuine multi-qubit entanglement (i.e. shared by all the involved
qubits). This make it possible to propose some appropriate
information-theoretic measure of such genuine multi-qubit entanglement \cite%
{GME}, which is one of the most central issues in quantum information theory
\cite{Wong,Hyperdeterminants,Mayer,GE,Gerardo}.

The paper is structured as follows. In Sec II, we introduce the
measure of local information based on the definition of optimal
fidelity. In Sec III and IV, we investigate the information
dynamics and demonstrate the perfect complementary behavior
through simple examples in two- and three-qubit systems. In Sec V,
we formulate the information complementarity relation based on
appropriate measures of local information and entanglement. In Sec
VI are discussions and conclusions.

\section{Optimal fidelity and local information}

We start by considering the information content in one qubit. Suppose Alice
get one qubit in the state $|\varphi \rangle $ from a random source $%
\{|\Omega \rangle ,|\Omega ^{\bot }\rangle \}$ with the probabilities $%
\{1/2,1/2\}$. Alice hold the qubit in state $|\varphi \rangle $, and can
exactly eliminate the uncertainty about the state preparation, i.e. the
information content in $|\varphi \rangle $ should be 1 bit. If Alice sent
the qubit to Bob through quantum channels, the qubit Bob received becomes $%
\rho =\xi (|\varphi \rangle \langle \varphi |)$. If $\rho =\frac{1}{2}(I+%
\vec{r}\cdot \vec{\sigma})$, with $\vec{r}$ the Bloch vector and the Pauli
operators $\vec{\sigma}=(\sigma _{x},\sigma _{y},\sigma _{z})$, is a mixed
state, Bob can only tell which of the two states that Alice sent to him with
some success probability, which results from the information loss through
the quantum channels. The success probability can be characterized by the
fidelity $F=\langle \varphi |\rho |\varphi \rangle $ \cite{Nielsen & Chuang}%
. However, we note that Bob can apply some kind of physical realizable local
strategy to maximize the success probability. Therefore, we can naturally
define the optimal fidelity as
\begin{equation}
F_{o}(\rho )=\underset{A\in SU\left( 2\right) }{\max }\langle \varphi |A\rho
A^{\dagger }|\varphi \rangle
\end{equation}%
Here unitary operation $A$ can be interpreted as some kind of local strategy
for Bob to maximize the success probability. After simple calculations, it
can be seen that $F_{o}(\rho )=\frac{1}{2}(1+|\vec{r}|)$. The optimal
fidelity ranges from $\frac{1}{2}$ to $1$, i.e. $\frac{1}{2}\leq F_{0}(\rho
)\leq 1$. Therefore, Bob hold the qubit in state $\rho $ can eliminate a
part of uncertainty about the state preparation. The information content in
the state $\rho $ can thus be characterized by the above success probability
in Eq.(1). In principle any monotone function of the success probability can
serve as a measure of information content in the state $\rho $. We introduce
a measure of local information based on the optimal fidelity as
\begin{equation}
I_{F}(\rho )=[2F_{o}(\rho )-1]^{2}
\end{equation}%
where $I_{F}(\rho )$ is normalized such at $I_{F}(\rho )=0$ for $F_{o}(\rho
)=\frac{1}{2}$ and $I_{F}(\rho )=1$ for $F_{o}(\rho )=1$. We will show that $%
I_{F}(\rho )$ defined above is equivalent to an operationally invariant
information measure \cite{Brukner and Zeilinger} and is a suitable measure
of local information in the situation we discuss here. For an $N$-qubit
quantum system $|\psi \rangle $, the total local information is
\begin{equation}
I_{l}^{total}=\sum\limits_{i=1}^{N}I_{F}(\rho _{i})
\end{equation}%
where $\rho _{i}=Tr_{{1,\cdots ,i-1,i+1,\cdots ,n}}(|\psi \rangle \langle
\psi |)$ is the reduced density operator of the $ith$ qubit.

\section{Two-qubit system}

We start out by investigating the information dynamics through simple
examples in a system of two qubits with interaction between qubits. The
interaction between qubits can be described as the following Hamiltonian in
the canonical form with three parameters $c_{1},c_{2},c_{3}$ \cite{Bennett
et al @2002&Nielsen@2003}:
\begin{equation}
\mathbf{H}=c_{1}\sigma _{x}^{1}\otimes \sigma _{x}^{2}+c_{2}\sigma
_{y}^{1}\otimes \sigma _{y}^{2}+c_{3}\sigma _{z}^{1}\otimes \sigma _{z}^{2}
\end{equation}
Here, for simplicity, the local evolutions have been neglected. The initial
state is set as $|\psi(0)\rangle=|\Omega\rangle\otimes|0\rangle$, where $%
|\Omega\rangle=\alpha|0\rangle+\beta|1\rangle$ is some prescribed pure
state. Due to the interaction between two qubits, the information will be
transferred between two qubits and can also be converted into the form of
nonlocal information.

\textit{Ising coupling.} We first consider the Ising interaction, i.e. the
coupling parameter $c_{1}=c$, $c_{2}=c_{3}=0$. After some time $t$, the
system becomes $|\psi(t)\rangle=exp(-itH)|\psi(0)\rangle=\alpha\cos{ct}%
|00\rangle -i\beta\sin{ct}|01\rangle+\beta\cos{ct}|10\rangle-i\alpha\sin{ct}%
|11\rangle$. Therefore, we can calculate the local optimal fidelity defined
in Eq.(1) $F_{0}(\rho_{i},t)=\{1+[1-|\alpha^2-\beta^2|\sin^2(2ct)]^{1/2}\}/2$%
, which will lead to the total local information as
\begin{equation}
I_{l}^{total}(t)=2[1-|\alpha^2-\beta^2|\sin^2(2ct)]
\end{equation}

On the other hand, we can obtain the entanglement contained in the state $%
|\psi(t)\rangle$ measured by $2-$tangle, which is the square of concurrence
\cite{ConC}. After some simple calculations, we can get
\begin{equation}
\tau_{12}(t)=|\alpha^2-\beta^2|\sin^2(2ct)
\end{equation}

We depict the dynamics behavior of local information and
entanglement in Fig.(a). It can be seen that these two quantities
exhibit perfect complementary behavior during the time evolution.
In fact, this can be easily verified from the above Eq.(5-6) by
deriving the complementarity relation that $I_{l}^{total}(t)+2\tau
_{12}(t)=2$. It is well known that entanglement is some kind of
nonlocal information. Our results in this simple example
demonstrate that based on some suitable definition of local
information (e.g. $I_{F}$ here), the nonlocal information is
directly related to some appropriate measure of entanglement. In
the following section, we can see that this viewpoint of
entanglement and nonlocal information is also applicable in
three-qubit systems and for arbitrary two-qubit system Hamiltonian
$H$.

\begin{figure}[tbh]
\includegraphics[scale=0.5]{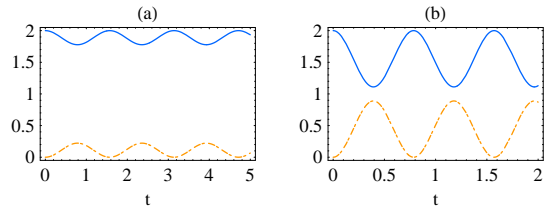}
\caption{(Color online) Information dynamics via Ising interaction (a) and
XY interaction (b). Total local information $I_{l}^{total}(t)$ \textit{vs.} $%
t$ (Solid) and nonlocal information $2\protect\tau_{12}(t)$ \textit{vs.} $t$
(Dashed). The coupling parameter $c=1$, the initial state is $|\protect\psi%
(0)\rangle=(\protect\sqrt{1/3}|0\rangle+\protect\sqrt{2/3}%
|1\rangle)\otimes|0\rangle$.}
\end{figure}

\textit{XY coupling.} In addition, we also demonstrate the quantum
state information dynamics for XY interaction, when the coupling
parameters of the Hamiltonian in Eq.(4) are $c_{1}=c_{2}=c,
c_{3}=0$. The result is depicted in Fig.1 (b), which exhibit the
same perfect complementary behavior as the situation of Ising
coupling.

\section{Three-qubit system}

In this section, we will extend the above discussions to three-qubit
systems. In such a system, not only two-qubit entanglement but also genuine
three-qubit entanglement, which is shared by all the three qubits.
Therefore, the corresponding relation between nonlocal information and
entanglement should be dealt with carefully. We consider a simple example
for demonstration, the system Hamiltonian is
\begin{equation}
H=\sum\limits_{ij}(c_{1}\sigma _{x}^{i}\otimes \sigma _{x}^{j}+c_{2}\sigma
_{y}^{i}\otimes \sigma _{y}^{j}+c_{3}\sigma _{z}^{i}\otimes \sigma _{z}^{j})
\end{equation}

The initial state is set as $|\psi(0)\rangle=|\Omega\rangle\otimes|0\rangle%
\otimes|0\rangle$, where
$|\Omega\rangle=\alpha|0\rangle+\beta|1\rangle$. We first
calculate the spectrum of the above Hamiltonian $H$, the eigen
energy is denoted as $\epsilon_{k}$ ($k=1,2, \cdots ,8$) and the
corresponding eigen state is $|\phi_{k}\rangle$. The initial state
can be expressed in the
eigen basis as $|\psi(0)\rangle=\sum\limits_{k=1}^{8}\gamma_{k}|\phi_{k}%
\rangle$. Therefore, the system state at time $t$ becomes $%
|\psi(t)\rangle=exp(-itH)|\psi(0)\rangle=\sum\limits_{k=1}^{8}%
\gamma_{k}e^{-i\epsilon_{k}t}|\phi_{k}\rangle$. According to Eq.(1-3), we
can obtain the total local information $I_{l}^{total}(t)$ easily. The
two-qubit entanglement between qubit $i,j$ ($ij=12,23,13$) measured by $2-$%
tangle $\tau_{ij}(t)$ can also be calculated from the reduced
density matrices $\rho_{ij}(t)$ \cite{ConC}. As we have mentioned
above, there will be another form of entanglement besides pairwise
entanglement, i.e. genuine three-qubit entanglement, in systems of
three qubits. The genuine three-qubit entanglement can be measured
by the $3-$tangle $\tau_{123}(t)$ proposed in \cite{CKW}. In order
to observe the perfect complementary behavior between the total
local information and entanglement, we should adopt some suitable
function which combine the contributions of both
two-qubit and three-qubit entanglement. Here we choose the function as $%
\mathcal{E}(t)=2[\tau_{12}(t)+\tau_{23}(t)+\tau_{13}(t)]+3\tau_{123}(t)$
and the information dynamics exhibit perfect complementary
behavior in Fig.(2). In the following section, we can see that the
coefficients before $\tau_{ij}$ and $\tau_{123}$ are not
arbitrary. In fact, this function has clear information-theoretic
meaning.

\begin{figure}[tbh]
\includegraphics[scale=0.5]{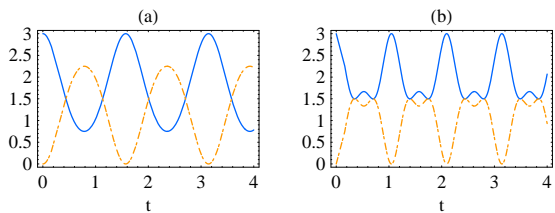}
\caption{(Color online) Information dynamics via Ising interaction (a) and
XY interaction (b). Total local information $I_{l}^{total}(t)$ \textit{vs.} $%
t$ (Solid) and nonlocal information $\mathcal{E}(t)$ \textit{vs.} $t$
(Dashed). The coupling parameter $c=1$, the initial state is $|\protect\psi%
(0)\rangle=(\protect\sqrt{1/3}|0\rangle+\protect\sqrt{2/3}%
|1\rangle)\otimes|0\rangle\otimes|0\rangle$.}
\end{figure}

Similar to the situation of two qubits, we can also easily derive
the complementarity relation as
$I_{l}^{total}(t)+\mathcal{E}(t)=3$. In three-qubit systems, both
two-qubit and three-qubit entanglement are nonlocal form of
information. The above complementary behavior imply that if we
choose suitable measures of local information and different levels
of entanglement (e.g. two-qubit and genuine three-qubit
entanglement here), the nonlocal information is just contributed
by the entanglement linearly with appropriate weights. Though we
only demonstrate this result via simple examples, we can see that
it is applicable for any Hamiltonian $H$ of three-qubit systems in
the following section.

\section{Information complementarity relation}

In this section, we will formalize the basic information
complementarity relation underline the above information dynamics.
We adopt the operationally invariant information measure proposed
by Brukner and Zeilinger \cite{Brukner and Zeilinger} as the
measure of local information, which is defined as the sum of
one-shot information gained over a complete set of mutually
complementary observables (MCO). Consider a pure n-qubit state, if
measured by operationally invariant information measure, the total
information content is $n$ bit and is completely contained in the
system. For a spin-$1/2$ system with the density matrix $\rho $,
the operationally invariant information content is
\begin{equation}
I_{BZ}(\rho )=2Tr\rho ^{2}-1.
\end{equation}%
Therefore, for an $n$-qubit quantum system in pure state $|\Omega \rangle $,
the amount of information in local form is $I_{local}(|\Omega \rangle
)=\sum\limits_{i=1}^{n}I_{i}$, where $I_{i}$ is the local information
measured by $I_{i}=I_{BZ}(\rho _{i})=\sum\limits_{i=1}^{n}(2Tr\rho
_{i}^{2}-1)$, where $\rho _{i}=Tr_{{1,\cdots ,i-1,i+1,\cdots ,n}}(|\Omega
\rangle \langle \Omega |)$ is the reduced density operator of the $ith$
qubit. The non-local information is $I_{non-local}=n-I_{local}$. We will
show that such non-local information is related to entanglement. In other
words, \textit{entanglement can be viewed as non-local form of information}.

We start by considering the simplest case of a two-qubit system in the pure
state $|\Omega\rangle_{12}=\sum\limits_{i,j=0,1}a_{ij}|ij\rangle$. The local
information contained in qubit $1$ and $2$ is $%
I_{1}=I_{2}=1-4|a_{00}a_{11}-a_{01}a_{10}|^{2}$. Therefore the non-local
information is $I_{non-local}=2-I_{1}-I_{2}=8|a_{00}a_{11}-a_{01}a_{10}|^{2}$%
. If measured by 2-tangle, which is the square of concurrence \cite{ConC},
the pairwise entanglement is $\tau_{12}=4|a_{00}a_{11}-a_{01}a_{10}|^{2}$.
Thus we can write local and non-local information as $I_{1}=I_{2}=1-\tau_{12}
$ and $I_{non-local}=2\tau_{12}$. The relation between local information and
nonlocal entanglement is depicted in Fig. 3 (A).

\begin{equation}
I_{1}+I_{2}+2\tau_{12}=2.
\end{equation}

\begin{figure}[tbh]
\includegraphics[scale=0.3]{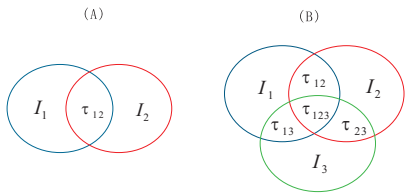}
\caption{(Color online) Information diagram for local information and
entanglement. (A). two-qubit pure states; (B) three-qubit pure states. One
circle represent one bit information.}
\end{figure}

If we focus on one qubit, say qubit $1$, then the $1$ bit information of
this qubit is partly contained in itself, \textit{i.e.}, $I_{1}$. The
residual information is contained in its entanglement with its environment,
in this case qubit $2$. The amount of this kind of information is $\tau_{12}$%
. It can be seen that $I_{1(2)}+\tau_{12}=1$. If the system is not isolated,
in general it will be in a mixed state $\rho_{12}$. In this case, the $2$
bit information is not only contained in the system but also in its
correlations with the outside environment. Therefore, $I_{1}+I_{2}+2%
\tau_{12}<2$ for a mixed state $\rho_{12}$. This result can be proved
through the convexity of $I_{1}$, $I_{2}$ and $\tau_{12}$.

Now we will extend our discussions to the case of three-qubit systems in a
pure state. The total information content in this system is $3$ bit. The
local information contained in each individual qubit is $I_{m}=2Tr%
\rho_{m}^{2}-1$, $m=1,2,3$, where $\rho_{m}$ is the reduced density matrix
for each qubit. Different from the two-qubit system, in this case the
non-local information exists not only in 2-qubit entanglement, but also in
genuine 3-qubit entanglement. It can be written as $%
I_{nonlocal}=3-(I_{1}+I_{2}+I_{3})$. Similar to the case of two-qubit, the
nonlocal information in two-qubit entanglement is $%
I_{nonlocal}(2)=I_{12}+I_{13}+I_{23}=2(\tau_{12}+\tau_{13}+\tau_{23})$,
where $\tau_{ij}$ is the 2-tangle between qubit $i$ and $j$. Therefore the
residual non-local information in the form of genuine three-qubit
entanglement should be $I_{nonlocal}(3)=3-(I_{1}+I_{2}+I_{3})-2(\tau_{12}+%
\tau_{13}+\tau_{23})$. We note that $2\tau_{12}=2(\lambda_{12}^{1}-%
\lambda_{12}^{2})^{2}=(1-I_{1}-I_{2}+I_{3})-4\lambda_{12}^{1}\lambda_{12}^{2}
$, where $\lambda_{12}^{1} \geq \lambda_{12}^{2}$ are the squared roots of
the eigenvalues of $\rho_{12}\tilde{\rho}_{12}$. Here $\tilde{\rho}_{12}=(%
\hat{\sigma}_{y}\bigotimes\hat{\sigma}_{y})\rho_{12}^{*}(\hat{\sigma}%
_{y}\bigotimes\hat{\sigma}_{y})$ is the time-reversed density matrix of $%
\rho_{12}$. Similarly $2\tau_{13}=(1-I_{1}-I_{3}+I_{2})-4\lambda_{13}^{1}%
\lambda_{13}^{2}$ and $2\tau_{23}=(1-I_{2}-I_{3}+I_{1})-4\lambda_{23}^{1}%
\lambda_{23}^{2}$. Then the residual non-local information in the form of
genuine three-qubit entanglement is $I_{nonlocal}(3)=4(\lambda_{12}^{1}%
\lambda_{12}^{2}
+\lambda_{13}^{1}\lambda_{13}^{2}+\lambda_{23}^{1}\lambda_{23}^{2})$. If
measured by 3-tangle proposed in \cite{CKW}, the genuine 3-qubit
entanglement is $\tau_{123}=4\lambda_{12}^{1}\lambda_{12}^{2}=4%
\lambda_{13}^{1}\lambda_{13}^{2}=4\lambda_{23}^{1}\lambda_{23}^{2}$.
Therefore, we can establish a direct relation between nonlocal information
and some appropriate measure of genuine three-qubit entanglement, i.e. $%
I_{nonlocal}(3)=3\tau_{123}$. The complementarity relation between
local information and entanglement is as follows
\begin{equation}
I_{1}+I_{2}+I_{3}+2(\tau_{12}+\tau_{13}+\tau_{23})+3\tau_{123}=3.
\end{equation}
Here, we present the complementarity relations in the formal
information-theoretic framework. If the system is in a mixed
state, the
above equation will be replaced by an inequality, \textit{i.e.}, $%
I_{1}+I_{2}+I_{3}+2(\tau_{12}+\tau_{13}+\tau_{23})+3\tau_{123}\leq3$. If we
focus on qubit $1$, then the $1$ bit information of this qubit is partly
contained in itself ($I_{1}$). The residual information is contained in its
entanglement with its environment, \textit{i.e.}, qubit $2$ and $3$. In fact
the relation $I_{1}+\tau_{12}+\tau_{13}+\tau_{123}=1$ is satisfied for qubit
$1$. Similar results hold for qubit $2$ and $3$. The above results are
depicted in Fig. 3 (B).

If we write the local reduced density matrix as $\rho =\frac{1}{2}(I+\vec{r}%
\cdot \vec{\sigma})$, with $\vec{r}$ the Bloch vector, the measure
of local information proposed by Brukner and Zeilinger as in
Eq.(8) $I_{BZ}(\rho )=2Tr\rho ^{2}-1=|\vec{r}|^{2}$. It can be
easily verified that this measure of local information is
equivalent to the measure based on optimal fidelity in Eq.(2),
i.e.,
\begin{equation}
I_{BZ}(\rho )=I_{F}(\rho )
\end{equation}
Therefore, the complementarity relations between local information
quantified by $I_{F}(\rho )$ and entanglement and that between
local information based on $I_{BZ}(\rho )$ and entanglement are
equivalent in nature. However, it can be seen from their
definitions that $I_{F}(\rho )$ and $I_{BZ}(\rho )$
have different physical meanings. The local information based on fidelity $%
I_{F}(\rho )$ is defined from the viewpoint of quantum
communications, while $I_{BZ}(\rho )$ is an operationally
information measure from the measurement viewpoint \cite{Brukner
and Zeilinger}. Since the above two complementarity relations are
equivalent, we could obtain that the complementarity relation
between local information quantified by $I_{F}(\rho )$ and
entanglement is hold for arbitrary initial pure states. In
particularly, the initial states could be entangled states, which
means that the isolated system contains nonlocal information
initially. Due to the interactions, the entanglement will change,
which results in the change of nonlocal information.

\section{Conclusions and discussions}

In summary, we adopt a measure of local information based on
optimal fidelity to investigate the information dynamics in two-
and three-qubit systems with interactions between qubits. Through
simple examples, we demonstrate the perfect complementary behavior
between local information and entanglement. We also show that the
measure of local information based on optimal fidelity is
equivalent to the operationally information measure proposed by
Brukner and Zeilinger. Furthermore, we establish a direct relation
between nonlocal information and different levels of entanglement,
and formalize the information complementarity relation by some
appropriate measures of local information and entanglement.

For two-qubit pure states, using von Neumann entropy as a measure
of local information, there has been a similar complementarity
relation between local information and entanglement, which is
measured by entanglement of formation \cite{Horodecki}. Here we
adopt a measure of local information by using linear entropy
rather than von Neumann entropy. This is based on the following
two considerations. One is that linearity always implies
additivity, which is simple and suitable for establishing
complementarity relations. The other point is that using linear
entropy we demonstrate that nonlocal information can be directly
related to the polynomial measures of entanglement, i.e. k-tangle.
For the situation of two qubits, 2-tangle is just the square of
concurrence, which is a function of entanglement of formation.
However, there are no straightforward generalization of
entanglement of formation to quantum states of more than two
qubits. Therefore, using linear entropy, it is more simple for us
to generalize the information complementarity relations to the
situations of more two qubits straightforwardly .

Though the relation between nonlocal information and entanglement
is demonstrated for two- and three-qubit pure states. It is
possible to generalize the information complementarity relation to
arbitrary $n$-qubit pure states naturally as the following
conjecture
\begin{equation}
\sum\limits_{i} I_{i}+2\sum\limits_{i_{1}<i_{2}}\tau_{i_{1}i_{2}}+ \cdots +
n \sum\limits_{i_{1}<i_{2}<\cdots <i_{n}}\tau_{i_{1}\cdots i_{n}}= n
\end{equation}
where $\tau_{i_{1}\cdots i_{k}}$ $(k=2,3,\cdots, n)$ are some
appropriate measures of genuine $k-$qubit entanglement. Since
nonlocal information is contributed by different levels of
entanglement as can be seen from the above discussions, conversely
we can characterize entanglement through nonlocal information. In
our recent work \cite{GME}, we have proposed such an
information-theoretic measure of genuine multi-qubit entanglement,
and utilize it to explore the genuine multi-qubit entanglement in
spin systems.

\section{Acknowledgment}
This work was funded by National Fundamental Research Program, the
Innovation funds from Chinese Academy of Sciences, NCET-04-0587,
and National Natural Science Foundation of China (Grant No.
60121503, 10574126).

\end{document}